\documentclass[12pt]{revtex4-1}
\usepackage{amsfonts,amsmath,amssymb,xcolor}
\usepackage{amsthm,amscd,ulem,bm}
\usepackage{graphicx}
\makeatletter

\@addtoreset{equation}{section}
\makeatother
\theoremstyle{definition}

\newcommand{\bib}[2]{\frac{\partial {#1}}{\partial {#2}}}
\newcommand{\bbib}[3]{\frac{\partial^2 {#1}}{\partial {#2}{\partial {#3}}}}

\def\b1{{ {\bf 1}}}

\begin{document}
\title{
Cusp singularity in mean field Ising model
}
%%
%\author{Yayoi Abe}
%\email{CQA11015@nifty.com}
%\affiliation{Physics Department, Ochanomizu University, 2-1-1 Ootsuka Bunkyo, Tokyo, Japan }
%\author{Muneyuki Ishida}
%\email{ishida@phys.meisei-u.ac.jp}
%\affiliation{Departmentof Physics, Meisei University,
% 2-1-1 Hodokubo, Hino, Tokyo 191-8506, Japan }
%\author{Erika Nozawa}
%\email{g1640613@edu.cc.ocha.ac.jp}
%\affiliation{Physics Department, Ochanomizu University, 2-1-1 Ootsuka Bunkyo, Tokyo, Japan }
%\author{Takayoshi Ootsuka}
%\email{ootsuka@cosmos.phys.ocha.ac.jp}
%\affiliation{Physics Department, Ochanomizu University, 2-1-1 Ootsuka Bunkyo, Tokyo, Japan }
%\author{Ryoko Yahagi}
%\email{yahagi@hep.phys.ocha.ac.jp}
% \affiliation{Physics Department, Ochanomizu University, 2-1-1 Ootsuka Bunkyo, Tokyo, Japan }

\author{Yayoi Abe$^a$, Muneyuki Ishida$^{b,\dagger}$, Erika Nozawa$^c$, Ryoko Yahagi$^d$\\ 
and\\ Takayoshi Ootsuka$^{e,\dagger\dagger}$\\
}
\email{$^a$ CQA11015@nifty.com\\
$^b$ ishida@phys.meisei-u.ac.jp\\
$^c$ g1640613@edu.cc.ocha.ac.jp\\
$^d$ yahagi@hep.phys.ocha.ac.jp\\
$^e$ ootsuka@cosmos.phys.ocha.ac.jp
}
\affiliation{Physics Department, Ochanomizu University, 2-1-1 Ootsuka Bunkyo, Tokyo, Japan\\
$^\dagger$ Departmentof Physics, Meisei University, 2-1-1 Hodokubo, Hino, Tokyo 191-8506, Japan \\
$^{\dagger\dagger}$NPO Gakujutsu-Kenkyu Network, Japan
}

\date{\today}

%%%%%%%%%%%%%%%%
\begin{abstract}
%%%%%%%%%%%%%%%%
An entropy of the Ising model in the mean field approximation is derived by
the Hamilton-Jacobi formalism.
We consider a grand canonical ensemble with respect to the temperature and
the external magnetic field.
A cusp arises at the critical point, which shows a simple and new 
geometrical aspect of this model.
In educational sense, this curve with a cusp helps students
acquire a more intuitive view on statistical phase transitions.

%%%%%%%%%%%%%%%%
\end{abstract}
%%%%%%%%%%%%%%%%

\maketitle

%%%%%%%%%%%%%%%%%%%%%%
\section{Introduction}
%%%%%%%%%%%%%%%%%%%%%%

Phase transitions, critical phenomena, and the corresponding critical exponents 
are fundamental topics in statistical mechanics.
Though they should have a close relation with 
critical points of maps~\cite{Milnor}, catastrophe theory~\cite{Thom}, 
and singularity theory~\cite{Arnold, Izumiya} in mathematics,
there is little application to these simple physical problems.
In these geometrical standpoints,
it is natural to expect critical points to be singularities
on a certain surface or a curve.
A critical phenomenon, accompanied by the Hamilton-Jacobi structure, is 
simply visualized in the present paper.
We take Ising model in the mean field approximation as an example.

Relations between thermodynamics and Hamilton-Jacobi theory have been
discussed for many years
~\cite{Cara, Snow, Hermann, Mrugala, Ruppeiner}.
However, they are mostly considered under the quasi-static conditions.
Among them, one notable proposal was offered by Suzuki~\cite{Suzuki}.
The second law of thermodynamics can be considered as a variational principle
that determines reversible or irreversible processes.
He recognized a Finsler structure in the variational principle,
and identified the equation of state, or the virial relation, as the constraint which
inevitably arises in the Finsler-Lagrangian formulation~\cite{Suzuki, Ootsuka1, OYIT},
and derived a Hamilton-Jacobi structure.
The idea was supplemented by one of the authors~\cite{Ootsuka2}.

Following Suzuki's method, we review the Finsler-Lagrangian formulation in the next section.
In section 3, we apply this method to the mean field Ising model
and show several graphs which reveal singularities at the critical points.

%%%%%%%%%%%%%%%%%%%%%%%%%%%%%%%%%%%%%%%%%%%%%%%%%%%%%%%%
\section{Finsler-Lagrangian 
formulation and Suzuki's Hamilton-Jacobi thermodynamics}
%%%%%%%%%%%%%%%%%%%%%%%%%%%%%%%%%%%%%%%%%%%%%%%%%%%%%%%%

\subsection{Review of Finsler-Lagrangian formulation}

Here, we review a Finsler geometrical formulation of Lagrangian formalism,
which we call Finsler-Lagrangian formulation
~\cite{Ootsuka1, OYIT, Ootsuka2, Lanczos, Erico}.
Let $Q=\{(q^1,q^2,\dots,q^n)\}$ be a configuration space and 
$L=L\left(q^i,\dot{q}^i,t\right)$ be a Lagrangian.
It is well known that the Lagrangian $L$ constructs a Finsler metric $F$
on the extended configuration space $M:=\mathbb{R}\times Q$ as
\begin{align}
 F=F\left(x^\mu,dx^\mu\right):=L\left(x^i,\frac{dx^i}{dx^0}, x^0 \right)dx^0.
 \label{LtoF}
\end{align}
The set $(M,F)$ becomes $(n+1)$-dimensional Finsler manifold.
This technique is known as homogenization technique.
In mathematics, Finsler metric should satisfy several conditions
~\cite{Matsumoto, Chern}, which are too strong for physical applications.
We only assume 1)~homogeneity of $F$ and 2)~domain of $F$ as
\begin{align}
 &
 1) \quad F(x,\lambda dx)=\lambda F(x,dx), \quad
 \lambda >0, \\
 &
 2) \quad F:D(F) \subset TM \to \mathbb{R},
\end{align}
where $D(F)$ is a subbundle of $TM$ where $F$ and its derivative is well defined.
Time evolutions of the system are
represented by oriented curves ${\cal C}=\{\bm{c}\}$
on the extended configuration space $M$.
The action of the Lagrangian system is defined 
by a line integral of $F=F(x,dx)$ along an oriented curve $\bm{c} \in {\cal C}$
\begin{align}
 {\cal A}[\bm{c}]=\int_{\bm{c}} F:=\int_{\tau_0}^{\tau_1} 
 F\left(x^\mu(\tau),\frac{dx^\mu}{d\tau}(\tau)\right)d\tau,
\end{align}
where $\tau$ is an arbitrary parameter of $\bm{c}$.
By homogeneity condition 1),
${\cal A}[\bm{c}]$ does not depend on the choice of the parametrization.
The variational principle
\begin{align}
 0=\delta {\cal A}[\bm{c}]=\int_{\tau_0}^{\tau_1} \left\{
 \bib{F}{x^\mu} \delta x^\mu+\bib{F}{dx^\mu} \frac{d\delta x^\mu}{d\tau}
 \right\} d\tau
 =\int_{\tau_0}^{\tau_1} \left\{
 \bib{F}{x^\mu}-\frac{d}{d\tau} \left(\bib{F}{dx^\mu}\right)
 \right\} \delta x^\mu d\tau,
\end{align}
leads to a covariant Euler-Lagrange equation
\begin{align}
 0=\bib{F}{x^\mu}-\frac{d}{d\tau}\left(\bib{F}{dx^\mu}\right),\label{EL-eq}
\end{align}
where
$\displaystyle \bib{F}{x^\mu}\left(x(\tau),\frac{dx}{d\tau}(\tau)\right)$ and
$\displaystyle \bib{F}{dx^\mu}\left(x(\tau),\frac{dx}{d\tau}(\tau)\right)$
are functions of $x^\mu(\tau)$ and $\displaystyle \frac{dx^\mu}{d\tau}(\tau)$.
The important fact is the covariant Euler-Lagrange equation \eqref{EL-eq} is
parametrization invariant.
The homogeneity condition 1) is equivalent to the Euler's formula
\begin{align}
 F=\bib{F}{dx^\mu} dx^\mu,
\end{align}
and differentiating it with respect to $dx^\nu$ on both sides, we have
\begin{align}
 0=\bbib{F}{dx^\mu}{dx^\nu}dx^\mu=\bib{p_\mu}{dx^\nu}dx^\mu
  =\bib{p_\nu}{dx^\mu}dx^\mu,\label{noLeg}
\end{align}
where we define a covariant conjugate momentum
\begin{align}
 p_\mu:=\bib{F}{dx^\mu}.
\end{align}
(\ref{noLeg}) indicates the matrix 
$\displaystyle \left(\bib{p_\nu}{dx^\mu}\right)$
does not have the inverse matrix.
The inverse function theorem promises that there exists at least one constraint
\begin{align}
 G(x,p)=0,
\end{align}
among the variables $(x^\mu,p_\mu)$.
%%%%%%%%%

When we consider a relativistic free particle on an $(n+1)$ dimensional Lorentzian manifold 
$(M,g)$, we can take a Finsler metric on $M$ as
\begin{align}
 F=mc\sqrt{g_{\mu\nu}(x)dx^\mu dx^\nu}.
\end{align}
In this case, we have a constraint
\begin{align}
 G(x,p)=g^{\mu\nu}(x)p_\mu p_\nu -(mc)^2=0.
\end{align}
For a non-relativistic particle under a potential force,
the corresponding Finsler metric is
\begin{align}
 F=\frac{m g_{ij}(x)dx^i dx^j}{2dx^0}-V(x)dx^0, \quad (i,j=1,2,3),
\end{align}
and we get 
\begin{align}
 G(x,p)=p_0+\frac{1}{2m}g^{ij}(x)p_ip_j+V(x)=0,
\end{align}
as a constraint. 
%%%%%%%%%

Hamilton's principal function $W$ is defined 
as a line integral along a solution curve of (\ref{EL-eq}),
\begin{align}
 W(\xi_1;\xi_0):=\int_{\bm{c}_{\xi_0}^{\xi_1}} F
 =\int_{\bm{c}_{\xi_0}^{\xi_1}} p_\mu dx^\mu, \quad
p_\mu=\bib{F}{dx^\mu}(x,dx),
\label{eq2-8}
\end{align}
where, $\bm{c}_{\xi_0}^{\xi_1}$ is a solution curve 
connecting between 
$\xi_0\in M$ and $\xi_1 \in M$.
When $\xi_0$ is fixed, $W$ can be considered as a function on $M$: 
$W(\ \  ;\xi_0):\xi_1\in M \mapsto \mathbb{R}$.
An infinitesimal transformation 
$\delta_{\varepsilon \bm{v}}=\varepsilon {\cal L}_{\bm{v}}, \,
\xi_1 \mapsto \xi_1+\varepsilon \bm{v}$ 
which acts only the neighborhood of $\xi_1$ leads to
\begin{align}
 W(\xi_1+\varepsilon \bm{v};\xi_0)-W(\xi_1;\xi_0)
 &%=\delta_{\bm{v}} W(\xi_1;\xi_0)
 =\varepsilon {\cal L}_{\bm{v}}W(\xi_1) 
  =\varepsilon \iota_{\bm{v}}dW(\xi_1)
 \notag \\
 =\int_{\bm{c}_{\xi_0}^{\xi_1+\varepsilon \bm{v}}}p_\mu dx^\mu
  -\int_{\bm{c}_{\xi_0}^{\xi_1}}p_\mu dx^\mu
 & =\varepsilon \iota_{\bm{v}}\bigl\{
 p_\mu\left(\dot{\bm{c}}_{\xi_1}\right) dx^\mu\bigr\}(\xi_1),
\end{align}
for arbitrary $\bm{v}$.
Here, $p_\mu(\dot{\bm{c}}_{\xi_1})$ stands for the quantity $p_\mu(x,dx)$
contracted by the velocity $\dot{\bm{c}}_{\xi}$ at $\xi_1$.
Therefore, the principal function $W=W(\ \ ;\xi_0)$ admits the relation
\begin{align}
 dW=p_\mu dx^\mu, \qquad p_\mu=\bib{W}{x^\mu}.
\end{align}
The constraint $G(x,p)=0$ becomes
\begin{align}
 G\left(x^\mu, \bib{W}{x^\mu}\right)=0.
\end{align}
This is the covariant expression of the Hamilton-Jacobi equation,
which is necessarily derived from the homogeneity condition 
1) in the Finsler-Lagrangian formulation.

%%%%%%%%%%%%%%%%%%%%%%%%%%%%%%%%%%%%%%%%%%%%%%%%%%%%%%%%%%%%%%%%%%
\subsection{Second law of thermodynamics as variational principle}
%%%%%%%%%%%%%%%%%%%%%%%%%%%%%%%%%%%%%%%%%%%%%%%%%%%%%%%%%%%%%%%%%%

Let $\delta Q$ be the quantity of heat flowing into the system from the environment
of temperature $T^{ex}$ during an infinitesimal process,
and $dS$ be the difference of entropy between the initial and final equilibrium states.
The second law of thermodynamics can be written as
\begin{align}
 dS \geqq \frac{\delta Q}{T^{ex}}.
\end{align}
If equality is satisfied, the thermal process is reversible.
On the other hand, inequality represents irreversible process.
We will assume the right-hand side is supposed to be given 
by an integration of 
some Finsler metric defined on
thermodynamic state space $M=\{(U,V)\}$~\cite{Ootsuka2}:
\begin{align}
 \int_{a \to b} \frac{\delta Q}{T^{ex}}=\int_{\bm{c}_a^b}F(U,V,dU,dV),
 \label{thermoFin}
\end{align}
where $\bm{c}_a^b$ is a thermal process which is represented by an oriented
curve on $M$.
Reversible processes maximize this integral.
Integral of $\displaystyle \frac{\delta Q}{T^{ex}}$ on a
reversible process $\bm{c}_a^b$ becomes the entropy difference between 
$a$ and $b$:
\begin{align}
 S(b)-S(a) = \int_{\bm{c}_a^b} \frac{\delta Q}{T^{ex}},
\end{align}
and the maximal (stationary) integral of the RHS of (\ref{thermoFin})
gives the Hamilton's principal function $W$.
Therefore the Hamilton's principal function in thermodynamics is identical to the
entropy function: $W=S$.
Thus, we get the relation
\begin{align}
 dW=p_U dU+p_V dV=dS=\frac1T dU+\frac{p}{T}dV, \label{dW=dS}
\end{align}
where $p_U$ and $p_V$ are conjugate momenta of $U$ and $V$,
and the third equality of (\ref{dW=dS}) is the first law of thermodynamics.
From the above equation (\ref{dW=dS}),
we can conclude the covariant
conjugate
momenta of $(U,V)$ are
\begin{align}
 (p_U,p_V) = \left(\frac{1}{T},\frac{p}{T}\right).
 \label{eq2-16}
\end{align}
The constraint from the Finsler-Lagrangian formulation $G(x,p)=0$ 
turns into an equation:
\begin{align}
 G\left(U,V,\frac{1}{T}, \frac{p}{T}\right)=0.
\end{align}
Suzuki found out that this is the virial relation in thermodynamic system.

In the case of the ideal gas, it has the internal energy $\displaystyle U=\frac32 NkT$ 
and the equation of state $pV=NkT$, where $N$ is the number of the gas particles,
$k$ the Boltzmann constant, $T$ temperature, $p$ pressure, $V$ volume of the gas.
Its virial relation is
\begin{align}
 U=\frac32 pV.
\end{align}
%%%%%%%%%
With (\ref{eq2-16}), it becomes
\begin{align}
G(U,V,p_U,p_V) = p_U U - \frac32 p_V V = 0.
\end{align}
From this virial equation and
\begin{align}
 p_U=\bib{S}{U}, \qquad p_V=\bib{S}{V},\label{pUSpVS}
\end{align}
we can derive following Hamilton-Jacobi
equation of the ideal gas:
\begin{align}
 U\bib{S}{U}-\frac{3V}2\bib{S}{V}=0.
 \label{ideal}
\end{align}
The solution of the partial differential equation 
(\ref{ideal}) gives the entropy of the ideal gas
\begin{align}
 S=S(U,V)=\frac32 r \log{U}+r\log{V}+S_0,
\end{align}
where $r$ and $S_0$ are constants.
Using (\ref{pUSpVS}), we also have
\begin{align}
 p_U=\frac1T=\frac{3r}2 \frac{1}{U}, \qquad 
 p_V=\frac{p}{T}=\frac{r}{V},
\end{align}
which reproduce the internal energy and state equation of the ideal gas.
It is believed that the definition of the ideal gas needs both relations.
However, the procedure of this section tells 
that only the virial relation is needed,
and through the Hamilton-Jacobi equation, the rest follows.

%%%%%%%%%%%%%%%%%%%%%%%%%%%%%%%%%%%%%%%%%%%%%%%%%%%%%%%%
\section{Ising model in mean field approximation}
%%%%%%%%%%%%%%%%%%%%%%%%%%%%%%%%%%%%%%%%%%%%%%%%%%%%%%%%

We apply the formulation reviewed in section 2 to a spin system.
The Hamiltonian of Ising model in the mean field approximation is expressed as
\begin{align}
 H = \frac{NJz m^2}{2} - Jzm \sum_{i=1}^N S_i,
\end{align}
when there is no magnetic field.
Here, $N$ is the total site number, $J$ the strength of the interaction,
and $z$ the coordination number.
$m$ is the expectation value of an Ising spin 
$S_i = \pm 1$,
which admits the self-consistent equation.
We start with the grand canonical ensemble
\begin{align}
 \Xi
 = \sum_{\text{configuration}}
 \exp \left[ -\beta \left( \frac{NJz m^2}{2} - Jzm \sum_{i=1}^N S_i \right)
 - \xi \sum_{i=1}^N S_i \right], \label{z}
\end{align}
where $\displaystyle \beta=\frac{1}{kT}$.
$\xi$ is a parameter related to the fluctuation of the total magnetization
$\displaystyle M=Nm=\left\langle \sum_{i=1}^N S_i\right\rangle$,
and is proportional to the external magnetic field $h$.
Throughout this section, we assume $\xi$ to be nonzero,
since an infinitesimally small magnetic field is necessary for the phase transition.
The grand Massieu function $\Psi = k \log \Xi$, which is a function of
$\beta$ and $\xi$, generates the magnetization $M$
and the internal energy $U$ as
\begin{align}
 M
 & = - \frac{1}{k} \bib{\Psi}{\xi}
 = N \tanh \left( \frac{\beta JzM}{N} - \xi \right), \label{sce} \\
 U
 & = - \frac{1}{k} \bib{\Psi}{\beta}
 = - \frac{JzM^2}{2 N}. \label{um}
\end{align}
The equation \eqref{sce} is the self-consistent equation
and \eqref{um} gives a relation between $U$ and $M$ which should be kept all the time.
The entropy is given by 
\begin{align}
 S = \Psi - \beta \bib{\Psi}{\beta} - \xi \bib{\Psi}{\xi}
 = \Psi + k \beta U + k \xi M, \label{entropy1}
\end{align}
which derives its total derivative as
\begin{align}
 dS = k \beta dU + k \xi dM. \label{dS}
\end{align}
It means thermodynamic state space for this spin system is $\{(U,M)\}$ and
the conjugate momenta $p_U$ and $p_M$ are
\begin{align}
 p_U = \bib{S}{U} = k \beta, \hspace{10mm}
 p_M = \bib{S}{M} = k \xi. \label{par}
\end{align}
Additionally, \eqref{dS} has an information on the relation between the parameter $\xi$
and the magnetic field $h$.
From thermodynamic prediction, the energy change should be given by
\begin{align}
 dU = T dS - h dM, \hspace{10mm} \text{or} \hspace{10mm} 
 dS = k \beta dU + k \beta h dM.
\end{align}
Thus, we have $\xi=\beta h$.

The self-consistent equation \eqref{sce} is a candidate for virial relation.
However, it should not contain the statistical quantity $N$ inherently,
since virial relation is a concept of thermodynamics.
From the relation \eqref{um}, we have
\begin{align}
 m = \frac{M}{N} = - \frac{2U}{JzM}, \label{mU}
\end{align}
which should take a value between $\pm 1$.
Substituting it into \eqref{sce} to get rid of $N$, we obtain
our virial equation for the mean field Ising model:
\begin{align}
 \beta \frac{2U}{M} + \xi = \tanh^{-1} \left( \frac{2U}{JzM} \right). \label{sce2}
\end{align}
It transforms into
\begin{align}
 \frac{2U}{kM} \bib{S}{U} + \frac{1}{k} \bib{S}{M}
 = \tanh^{-1} \left( \frac{2U}{JzM} \right), \label{hj}
\end{align}
after substituting the derivatives of $S$ for $(\beta, \xi)$ in (\ref{sce2}) using (\ref{par}).
This is the Hamilton-Jacobi equation for the Ising model in mean field approximation.

\begin{figure}
 \includegraphics[width=80mm]{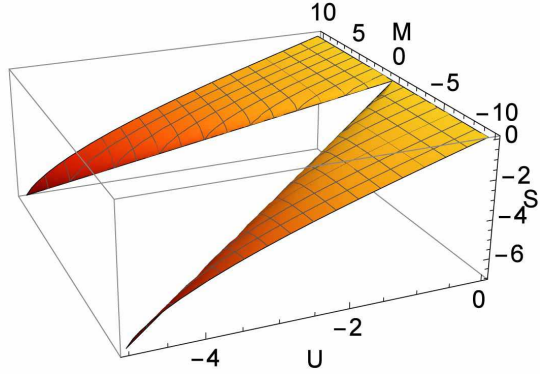}
 \caption{Entropy as a function of $U$ and $M$ ($k=Jz=1$).}
 \label{ent}
\end{figure}

\begin{figure}
 \begin{minipage}{0.4\hsize}
  \includegraphics[width=60mm]{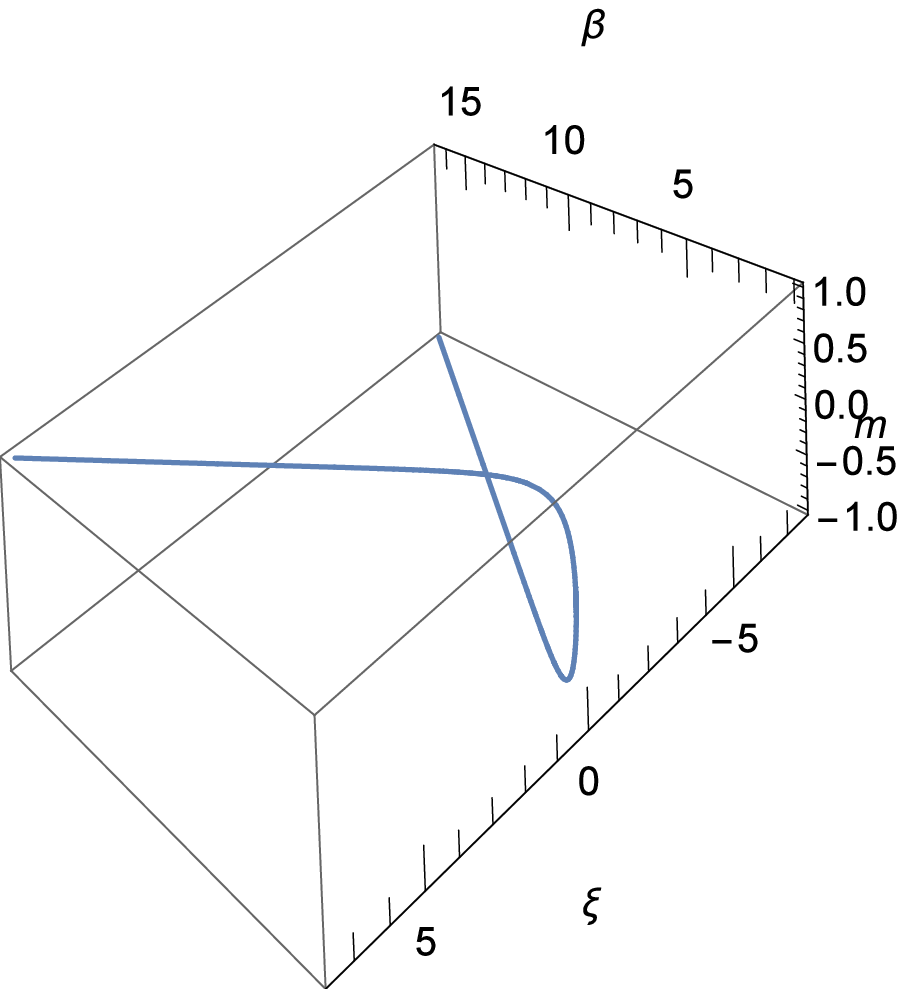}
  \caption{Solution curve in $(\beta, \xi, m)$.}
  \label{bxm}
 \end{minipage}
 \begin{minipage}{0.4\hsize}
  \includegraphics[width=60mm]{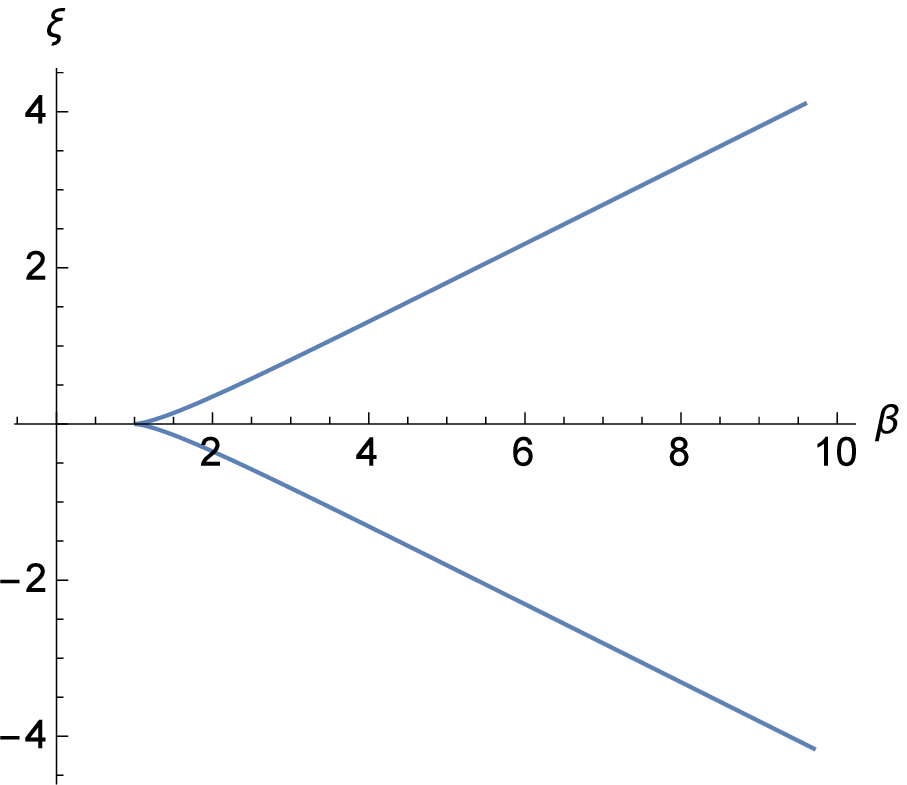}
  \caption{Graph of $(\beta, \xi)$.}
  \label{bx}
 \end{minipage}
\end{figure}

\begin{figure}
 \includegraphics[width=50mm]{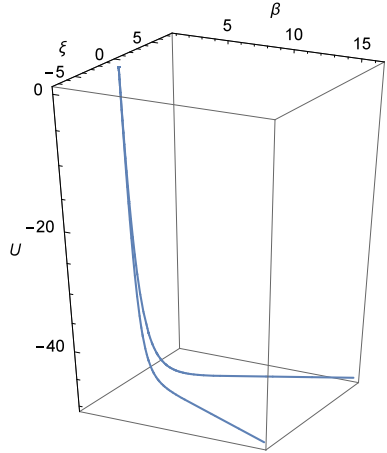}
 \caption{Graph of $(\beta, \xi, U)$.}
 \label{energy}
\end{figure}

\begin{figure}
 \begin{minipage}{0.4\hsize}
  \includegraphics[width=60mm]{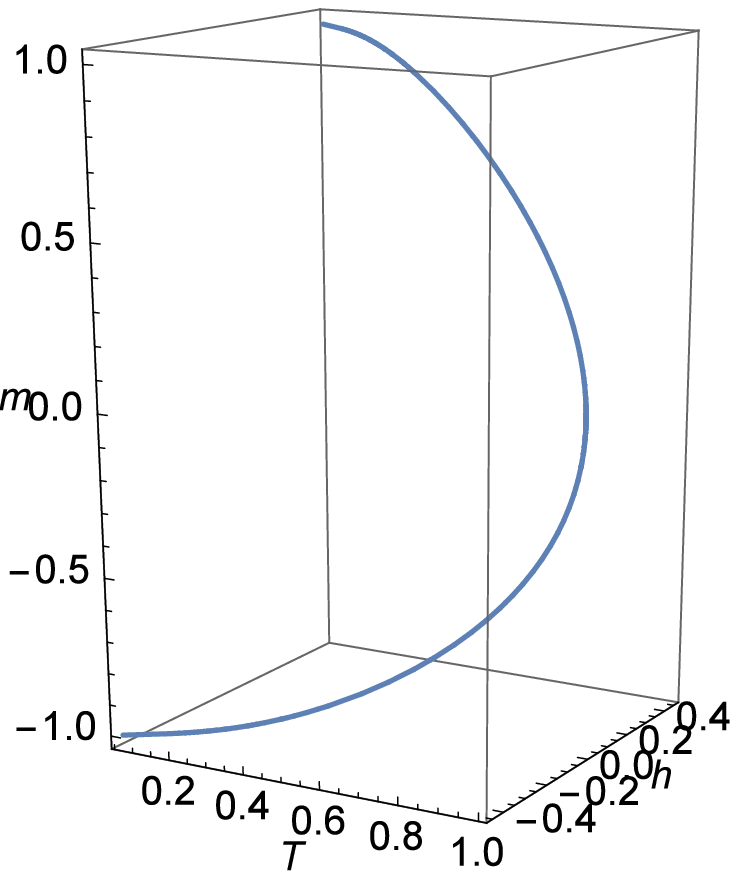}
  \caption{Solution curve in $(T,h, m)$.}
  \label{thm}
 \end{minipage}
 \begin{minipage}{0.4\hsize}
  \includegraphics[width=60mm]{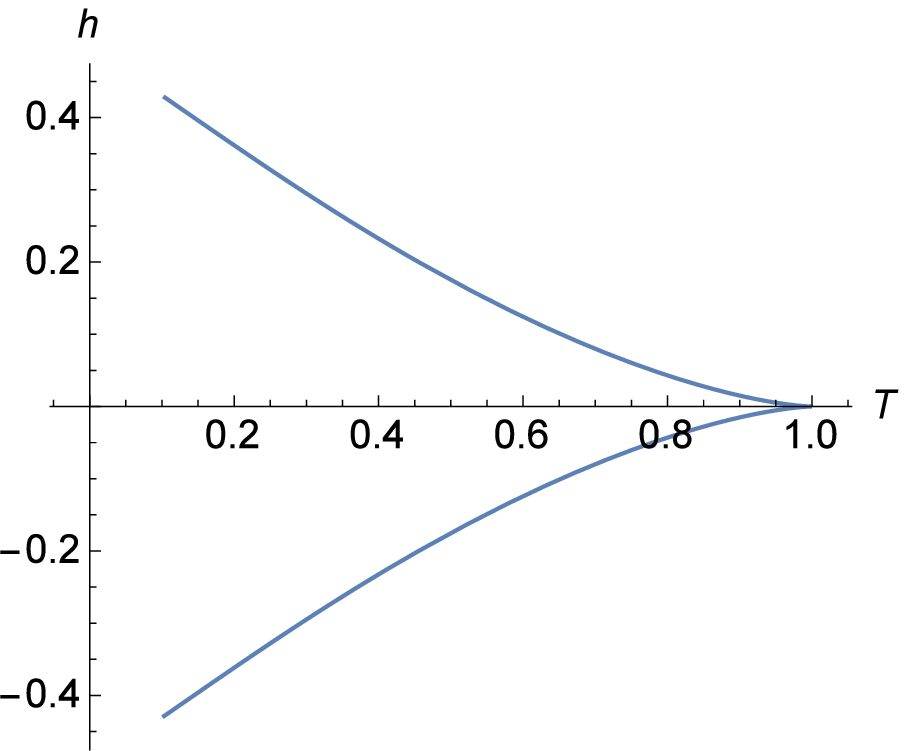}
  \caption{Graph of $(T,h)$.}
  \label{th}
 \end{minipage}
\end{figure}

\begin{figure}
 \includegraphics[width=50mm]{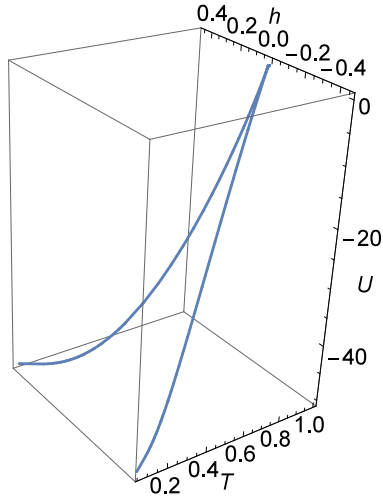}
 \caption{Graph of $(T,h, U)$.}
 \label{thu}
\end{figure}

By solving the partial differential equation \eqref{hj} directly,
assuming homogeneity of $S$ with respect to $(U,M)$,
we find the general solution for the entropy as
\begin{align}
 S
 = k M \tanh^{-1} \left( \frac{2U}{JzM} \right)
 + \frac{kJzM^2}{4U} \log \left( 1 - \left( \frac{2U}{JzM} \right)^2 \right)
 + a \frac{M^2}{U}, \label{entropy2}
\end{align}
where $a$ is an arbitrary constant.
The last term is set to be a linear function of $\displaystyle \frac{M^2}{U}$ 
because of the extensive property of the entropy.
When $U$ takes a value 0, this term pushes the entropy out to infinity,
which makes it unphysical.
Therefore we choose $a=0$ for a physical solution.
Its $(U,M)$-dependence is illustrated in FIG. \ref{ent}.
The vacant region in the middle indicates the non-allowed combination of $U$ and $M$
to be an argument of $\tanh^{-1}$.
After evaluating \eqref{par} and substituting \eqref{um}, we obtain
\begin{align}
 \beta
 & = - \frac{JzM^2}{4U^2}
 \log \left( 1 - \left( \frac{2U}{JzM} \right)^2 \right)
 = - \frac{1}{Jzm^2} \log \left( 1 - m^2 \right), \label{beta} \\
 \xi
 & = \tanh^{-1} \left( \frac{2U}{JzM} \right)
 + \frac{JzM}{2U} \log \left( 1 - \left( \frac{2U}{JzM} \right)^2 \right)
 = - \tanh^{-1} m - \frac{1}{m} \log \left( 1 - m^2 \right). \label{xi}
\end{align}
By substituting the above relations, the entropy \eqref{entropy2} becomes
\begin{align}
 S = - k N m \tanh^{-1} m - \frac{kN}{2} \log \left( 1 - m^2 \right), \label{entropy3}
\end{align}
which is identical to what is derived from \eqref{entropy1} except for a constant term.
The expression \eqref{entropy2}, or \eqref{beta}-\eqref{xi}, 
has much more information than \eqref{entropy3},
since $\beta$ and $\xi$ are parametlized by $m$.
It produces a curve in $\{(\beta, \xi, m)\}$ space as shown in FIG. \ref{bxm},
which resembles a partial curve of 
the famous cusp catastrophe surface.
FIG. \ref{bx} shows the projection of FIG. \ref{bxm} onto $\{(\beta, \xi)\}$ plane,
and we observe a cusp exactly at the critical point $kT_c=Jz$ 
and $h=0$ ($\beta=1$ for $k=Jz=1$).
Since the energy $U$ is also considered as a function of $m$, we can exhibit
the combination $(\beta, \xi, U)$ simultaneously (FIG. \ref{energy})
to see a drastic change in the energy at the critical point.
This fact gives a simple and new aspect of critical phenomenon as the singularity theory.
Graphs illustrated in terms of $(T,h)$ instead of $(\beta,\xi)$ are displayed in
FIGs \ref{thm}-\ref{thu}.

Geometrically thinking, the critical exponents should be determined 
along the path to the critical point in this $\{(\beta, \xi, m)\}$ space.
We verify that the curve in FIG. \ref{bxm} is the path 
for the mean field approximation.
The equations \eqref{beta} and \eqref{xi} expand as
\begin{align}
 \beta
 & = \frac{1}{Jz} + \frac{m^2}{2Jz} + \frac{m^4}{3Jz} + \cdots, \\
 \xi
 & = \frac{m^3}{6} + \frac{2m^5}{15} + \cdots.
\end{align}
From the second equation, we have $\delta=3$ for $m \propto |h|^{\frac{1}{\delta}}$.
The reduced temperature $t$ behaves as 
\begin{align}
 -t := \frac{T_c-T}{T_c}
 = \frac{Jz\beta-1}{Jz\beta}
 \simeq \frac{m^2}{2}, \label{t}
\end{align}
so that it gives the exponent $\displaystyle \beta=\frac{1}{2}$ for $m \propto |t|^\beta$. 
The differential 
$\displaystyle
\frac{d\xi}{dm}=\frac{d\beta}{dm}h+\beta
\frac{dh}{dm}$ 
leads the magnetic susceptibility $\chi$ to
\begin{align}
 \chi
 = \left. \frac{dm}{dh} \right|_{h=0}
 = \frac{\beta m^2(1-m^2)}{m^2 + (1-m^2) \log (1-m^2)} 
 \simeq \frac{\beta}{1-Jz\beta} 
 = \frac{1}{kT-kT_c}.
\end{align}
Thus we have $\gamma=1$ for $\chi \propto |t|^{-\gamma}$.
The specific heat $C$ becomes
\begin{align}
 C
 =\frac{dU}{dT}
 = - \frac{JzN}{2}\frac{d(m^2)}{dT}
 \simeq -JzN \frac{d}{dT} 
 \left( \frac{T_c-T}{T_c} \right)
 = kN,
\end{align}
by substituting \eqref{um} and \eqref{t}, and we get $\alpha=0$
for $C \propto |t|^{-\alpha}$.
All these exponents are the same as the standard results.

An exact solution for the higher-dimensional Ising model with non-zero magnetic field
would define an exotic surface as the cusp catastrophe surface
in $\{(\beta,\xi,m)\}$ space by considering the equation 
$\displaystyle m=-\frac{1}{kN}\bib{\Psi}{\xi}$.
However, it generally cannot define the unique exponents
since there are infinite numbers of paths approaching the critical point.
In contrast, the mean field Ising model gives a curve as seen in FIG. \ref{bxm},
so that we can define the unique exponents.

\begin{figure}
 \includegraphics[width=50mm]{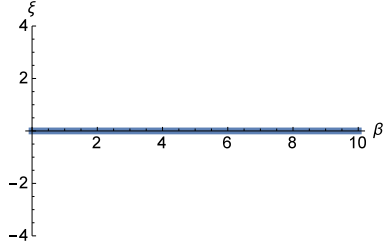}
 \caption{Graph of $(\beta, \xi)$ for $h=0$.}
 \label{bx2}
\end{figure}

The external magnetic field $h$, or $\xi$, is set to be nonzero throughout
the section 3, since this assumption is required to make the critical phenomenon happen.
Here, we remark on a mathematical solution of the self-consistent equation \eqref{sce}
for $h=0$ : $m=\tanh(\beta J z m)$.
The corresponding entropy satisfies the relation 
$\displaystyle \bib{S}{M}=0$,
which gives $S=\lambda U$ for some constant $\lambda$ due to its extensive property.
For $\beta J z \leq 1$, the only solution of the self-consistent equation is $m=0$.
We have $S=0$ from the relation \eqref{entropy3}.
If $\beta J z>1$, there exist $m \neq 0$ solutions and the entropy has a nonzero value.
Compared to the relation \eqref{par}, we have $\lambda=k\beta$.
Therefore the parameter $\lambda$ is restricted by 
$\displaystyle \lambda > \frac{k}{Jz}$.
FIG.\ref{bx2} shows its graphical description.
Though it is not an interesting solution, it obviously expresses a physical situation
with no magnetic field.

%%%%%%%%%%%%%%%%%%%%
\section{Discussion}
%%%%%%%%%%%%%%%%%%%%

We calculate the entropy of the Ising model in the mean field approximation
as a Hamilton's principal function on thermodynamic state space $\{(U,M)\}$.
Despite the fact that the Ising model is a statistical model,
this formalism push the site number $N$ away from the last results.
It extracts the thermodynamical state, which has
a clear singularity at the critical point.
The critical exponents are uniquely determined along the solution curve.
Standard calculations in various textbooks unknowingly assume this curve to
derive the correct exponents.

The solution curves, depicted in FIGS.1-4, are also considered to be a map $m\mapsto(\beta(m),\xi(m))$.
The rank of the corresponding Jacobi matrix goes to zero in $m\to0$ limit.
Thus $m=0$ is a critical point in terms of the singularity theory.
$\displaystyle 
(\beta,\xi)=\left(\frac{1}{kT_c},0\right)$
is the critical value.
It suggests that physical critical phenomena 
can be studied by the singularity theory,
and it also helps us acquire a more intuitive and simple geometric view.

%%%%%%%%%%%%%%%%%%%%%%%%%%%
\begin{acknowledgements}
%%%%%%%%%%%%%%%%%%%%%%%%%%%

We thank Prof. E. Tanaka, Prof. M. Morikawa and the astrophysics 
and cosmology lab (Ochanomizu university) for support and encouragements.

%%%%%%%%%%%%%%%%%%%%%%
\end{acknowledgements}
%%%%%%%%%%%%%%%%%%%%%%

%%%%%%%%%%%%%%%%%%%%%%%%%%%

%%%%%%%%%%%%%%%%%%%%%

%%%%%%%%%%%%%%%%
%%%%%%%%%%%%%%%%
%%%%%%%%%%%%%%%%
%%%%%%%%%%%%%%%%
\end{document}